# Nanolayer thickness detection via spatial mode projection

N. Hermosa[1,a)], C. Rosales-Guzmán[1], S. F. Pereira[2,b)], and J. P. Torres[1,3]

[1]ICFO- Institut de Ciencies Fotoniques, UPC, Mediterranean Technology Park, 08860 Castelldefels (Barcelona), Spain

[2]Optics Research Group, Delft University of Technology, Lorentzweg 1, Delft 2628 CJ, The Netherlands

[3]Department of Signal Theory and Communications, Universitat Politecnica de Catalunya, 08860, Barcelona, Spain

We demonstrate an optical scheme for measuring the thickness of thin nanolayers with the use of light beam´s spatial modes. The novelty in our scheme is the projection of the beam reflected by the sample onto a properly-tailored spatial mode. In the experiment described below, we are able to measure a step height smaller than 10 nm, i.e., one-eightieth $(1/80)$ of the wavelength with a standard error in the picometer scale. Since our scheme enhances the signal-to-noise ratio (SNR), which effectively increases the sensitivity of detection, the extension of this technique to the detection of subnanometric layer thicknesses is feasible.

The search for new optical methods to measure thickness in the range of a few nanometers or even hundreds of picometers is a topic of great interest. This is fuelled not only by the desire to reach the limit of resolution on the use of light in the nanoworld but also to develop new methods that can complement and/or substitute some well-established techniques such as X-ray spectroscopy, atomic force microscopy and ellipsometry [1,2,3]. Moreover, the continuous shrinking of all kinds of optical and electronic devices and the explosive growth of the exploration of the inner working of cells and molecular bio machines demand detection techniques that apart from being highly sensitive, must also be non-invasive, faster and easy to implement in different scenarios. These requirements can be met by photonics technologies.

Most of the time, high-resolution optical metrology is closely related to the evaluation of the phase of an electric field. In general, phases cannot be readily obtained and the desired information must be extracted indirectly by some other methods. The most widely used of these methods is interferometry. By looking at the

---

[a)]Author to whom correspondence should be addressed. Electronic mail: nathaniel.hermosa@icfo.es
[b)]This research was performed while S. F. Pereira was at the ICFO- Institut de Ciencies Fotoniqies, UPC, Mediterranean Technology Park, 08860 Castelldefels (Barcelona), Spain.

intensity produced at the output port of an interferometer, the relative phase can be measured and consequently, the relative thickness of a layer. Hugely small global phase differences between two independent beams up to $\sim 1\times 10^{-7}$ rad can be detected [4]. The detection of small structures, such as a step[5], is more cumbersome since the reflected beam contains a spatially-varying phase instead of a global phase which should be resolved.

A major problem in interferometry is the presence of uncontrollable disturbances that can also introduce phase differences. This is especially critical when tiny phase changes are being measured. A way to circumvent this problem is by using a common path interferometer (CPI) [6,7] where an unperturbed part of the beam acts as a reference beam and travels the same path as the signal beam. CPI has been used extensively in quantitative phase measurements since Dyson's seminal paper in 1953[8]. A CPI scheme at quadrature condition (i.e, the phase difference between the reference and signal beams is centered around $\pi/2$) is very sensitive to minute changes in the phase of the signal beam [9,10]. At this condition, CPI provides a linear relationship between the observed intensity modulation and the change in the optical phase that induces the intensity change.

One successful application of CPI at quadrature condition is in spinning-disk interferometry (SDI) [9,10]. SDI is used primarily in micro immunoassay wherein specific antigens attach to engineered substrates that fulfil the quadrature condition. Unfortunately, the need to fulfil the quadrature condition may limit the use of SDI since the phase ultimately depends on the wavelength, the thickness and the index of refraction of the substrate. Moreover there are situations where the quadrature condition cannot be achieved easily, such as when the required wavelength is either not available or might damage the sample.

In this letter, we put forward a novel way to circumvent this limitation. The key point of our approach is to project the reflected signal onto appropriately tailored spatial modes (*spatial mode projection*) before its power is measured. Instead of engineering surfaces or constructing spokes or ridges such that the quadrature condition is met that condition is passed on to the mode projection detection system. We demonstrate this method in an experiment where we measured a step height that is $1/80$ of the wavelength with a standard error in the range of picometers. We also show that the power of the beam upon projection has a linear dependence with tiny height changes for any substrate with an appropriate mode. Moreover by means of spatial mode projection, the signal-to-noise (SNR) is improved compared to standard SDI. We briefly mention here that there are other optical methods which use the difference in total power detected between light that are reflected with and without the sample to infer the thickness and refractive index of multilayers[11]. However, these methods are highly sensitive to the optical characteristics of the substrate and these do not work for low-loss highly reflecting samples.

In our scheme, the detected signal $P$ can be written as

$$P = \eta \int r(x,y) E_i(x,y) U(x,y) dxdy \tag{1}$$

where $\eta$ is the detection efficiency, $r(x,y)$ is the reflection coefficient, $E_i(x,y)$ is the incident electric field and $U(x,y)$ is the spatial mode we project onto. This projection can be efficiently made with diverse optical devices such as computer-generated holograms in spatial light modulators or liquid crystal switchable plates. The selection of mode to project onto the reflected light depends on the geometry of the sample. For simplicity, but without any loss of generality, we use a step sample in all our experiments. We define a step as a sharp discontinuity of height $h$ of the reflecting surface.

When a Gaussian mode illuminates a step, the most appropriate modes for projection are also Gaussian modes of the form

$$\begin{aligned} U(x,y) &\sim \exp\left(-\frac{x^2+y^2}{w_0^2}\right) & x \leq 0 \\ U(x,y) &\sim \exp\left(-\frac{x^2+y^2}{w_0^2}\right)\exp(i\Delta\varphi) & x > 0 \end{aligned} \tag{2}$$

where $w_0$ is the waist of the beam and $\Delta\varphi$ is the relative phase difference between the two regions of the modes with which we project onto. When a Gaussian beam reflects from this sample and is projected on the mode given by Eq. (2), the normalized detected intensity is given by

$$P_{-\Delta\varphi} = \frac{1}{2}\left(1+\cos(\delta+\Delta\varphi)\right) \tag{3}$$

where $\delta$ is related to the height of the cliff as $\delta = 4\pi h/\lambda$. Note that Eq. (3) is derived when the discontinuity in the phase of the mode coincides with the position of the cliff. We use Eq. (3) to measure samples with different step heights.

The experimental setup is shown in Fig. 1. A collimated HeNe laser ($\lambda = 632.8$ nm, $w_0 = 1.1$ mm) is incident on a sample with a step height of $h$. Our samples are etched SiO$_2$ on top of a Si wafer. We image the reflected light with a telescope onto the Spatial Light Modulator (SLM, Hamamatsu). The SLM is controlled to have a phase profile as in Eq (2). This is done by using appropriate 8-bit gray level values. Each gray level corresponds to a particular value of phase that the beam will acquire. Half of the beam acquires a phase of $\Delta\varphi$ with respect to its other half. The beam after the SLM is then sent to the single mode fiber (SMF). The SMF is connected to a Si photodector that is attached to a lock-in amplifier system. The laser beam is chopped before the sample and its chopping frequency is used as the reference frequency of the lock-in amplifier. The chopping frequency used is 3 kHz and the integration time of the lock-in amplifier is 300 ms. The data are logged by a digital oscilloscope (DO, Agilent) to the computer. In our measurements, we have also obtained the normalized intensity $P_{-\Delta\varphi}$, the power reflected for a phase $-\Delta\varphi$. This is done by switching the gray level values of the two regions of the SLM.

Figure 2 shows our experimental results for different step heights. Note the good correspondence of the experimental curve with the theoretical calculations. The values used for the theoretical curves are 0 nm, 8 nm and 31 nm, respectively. These are measured by a commercial profilometer (Alpha-Step IQ Surface profilometer). To quantify the height of the step from our measurements, we take the difference between $P_{\Delta\varphi}$ and $P_{-\Delta\varphi}$. The difference of the normalized powers is given by

$$P_{-\Delta\varphi} - P_{\Delta\varphi} = \sin\delta \sin\Delta\varphi. \qquad (4)$$

Note that the step height is readily accessible with $\sin\delta$ as $P_{-\Delta\varphi} - P_{\Delta\varphi}$ is plotted as a function of $\sin\Delta\varphi$. The uncertainty in the measured height comes from the standard error of the slope of the plotted line.

Figure 3 is an example of the analysis done to the experimental data. The experimental results fit nicely to the data. No fitting parameters are used. The maximum amplitude happens at quadrature $\Delta\varphi = \pi/2$ as seen in Fig. 3(a). The measured thickness layers are summarized in Table I. The uncertainty comes from the fact that the sample is not smooth as observed in the profilometer scans (*not shown*). Moreover, we ascribe the 1.9 nm offset in our data to the existence of a nonlinear relationship between the gray level value and the phase introduced by the SLM near $\pi$ which we found during initial calibration of the SLM (see for example the difference in the line fit with the theoretical line in Fig 3(b) near $\pi$ ).

As in the SDI, the maximum sensitivity in our scheme happens at the quadrature condition. The main difference, however, is the detection scheme. Consider for example a step geometry that fulfils the quadrature condition with a thin layer sample placed on top of the step. The system is illuminated by a Gaussian beam with beam waist $w_0$ and power $P_0$. The signal of interest is the normalized differential signal $P_1 - P_2$ where $P_1$ and $P_2$ are the power of the detected signal from the step with and without the thin layer. In the SDI, there is an optimum area of detection that gives the maximum value of $P_1 - P_2$. Increasing the area of detection decreases $P_1 - P_2$. In the mode projection scheme on the other hand, the total power of the projected signal is measured and hence, the power does not depend on the detection area. More importantly, given the proper choice of mode the signal-to-noise ratio will be higher. In the step geometry considering shot-noise condition, the SNR ratio is enhanced by ~ 2 dB compared to SDI when using spatial mode projection with $\Delta \varphi = 0$.

The importance of the quadrature condition is the large linear change in the differential signal produced by a tiny layer if the substrate is at quadrature. When the system is not at quadrature, the differential signal is diminished dramatically. This is not an issue in the scheme we present here as spatial modes can be easily engineered such that the linear dependence of the normalized differential signal $P_1 - P_2$ is preserved for any optical height $h/\lambda$. We show this for a thin layer on top of a flat substrate (inset Fig 4). In this case, the SDI will not work since the thin layer is on top of a substrate that does not meet the quadrature condition. In our scheme we can re-introduce the quadrature condition on the mode. Fig. 4 shows the differential signal when the reflected signals are projected onto Gaussian modes with different phase steps. Notice that at $\Delta \varphi = \pi/2$, the differential signal is linear with the height of the layer at the same time giving the maximum differential signal. On the contrary, $\Delta \varphi = \pi/4$ and $\Delta \varphi = 0$ do not give optimal differential signals. This is what we observe in our experiment (Fig. 3(a)). We note however that similar to SDI, the lateral resolution of our technique is diffraction limited.

The measurement of subnanometric steps requires the detection of small power differences $P_1 - P_2$. In our scheme, a normalized height $h/\lambda \sim 10^{-4}$ and an initial laser power of 1 mW would give a differential signal of $\delta/2 \sim 600\,\text{nW}$. This differential signal can be detected in principle. One can use the intensity measurement method by Freudiger et al.[12], where a high frequency detection scheme is used to get rid of lower-frequency laser noise, thus allowing the detection of a fractional power losses of up to $\sim 10^{-7}$. The signal–to–noise ratio is $S/N = \delta/2\sqrt{2\pi E\lambda/2\hbar c} \sim 14\,\text{dB}$ in our scheme using the parameters above.

In conclusion, we have demonstrated that extremely small step heights can be measured without the need to impose stringent conditions on the substrate by using spatial mode projection in a common path interferometer. We have measured a layer thickness as low as 9.7 nm with a standard error of 170 pm in our experiment. Moreover, we have shown that our scheme enhances the sensitivity of detection and hence, can also be used for subnanometric step height measurements.

This work was supported by the Government of Spain (project FIS2010-14831 and the program Excelencia Severo Ochoa), the European project PHORBITECH (FET-Open grant number: 255914), and the Fundacio Privada Cellex Barcelona. S.F.P. is grateful to ICFO for hosting her 6-month stay. Johan Osmond did the profilometry measurements.

FIGURES and TABLE

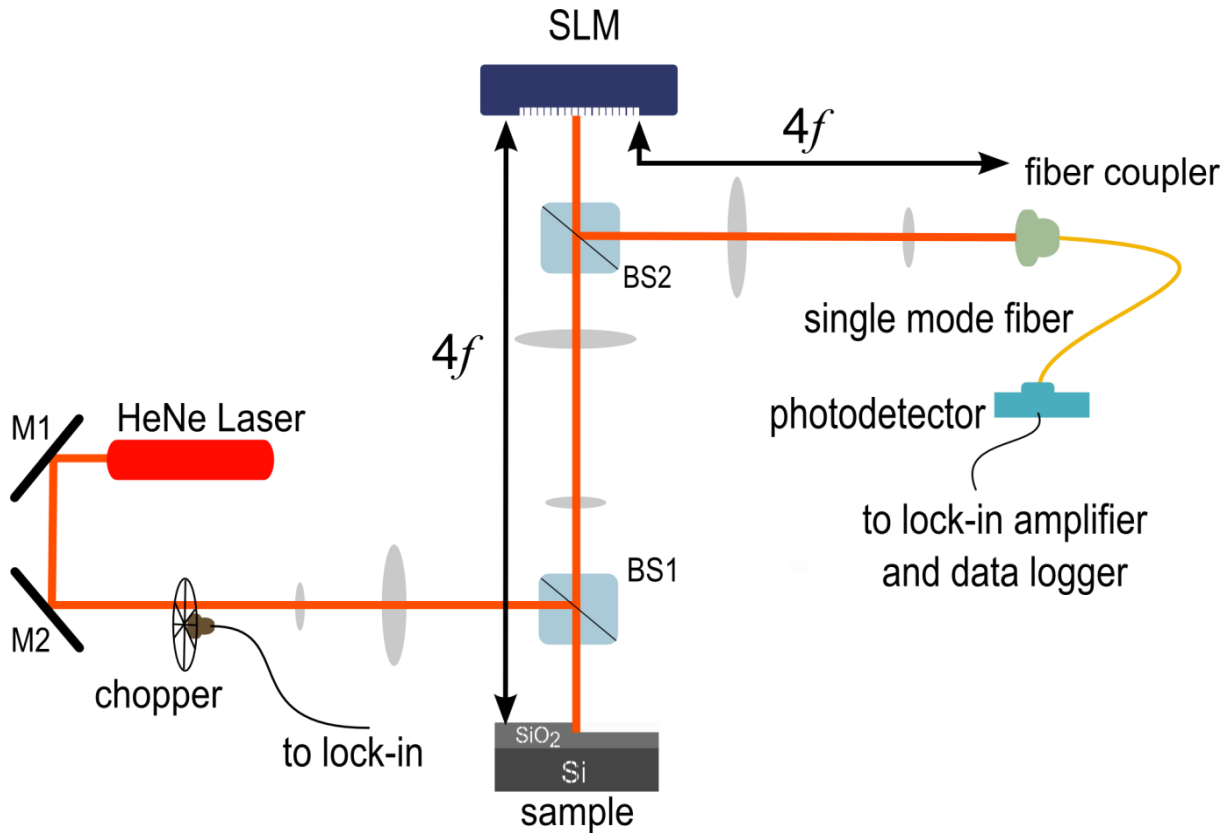

FIG 1. (Color online). The experimental setup. A HeNe laser beam impinges perpendicularly to the sample (the diagram of the sample is exaggerated). The reflection from the sample is then projected onto an SLM where a desired phase is encoded. The resulting beam is sent to a single mode fiber. A lockin amplifier with a chopper is used to lessen technical noise. *See text for details.*

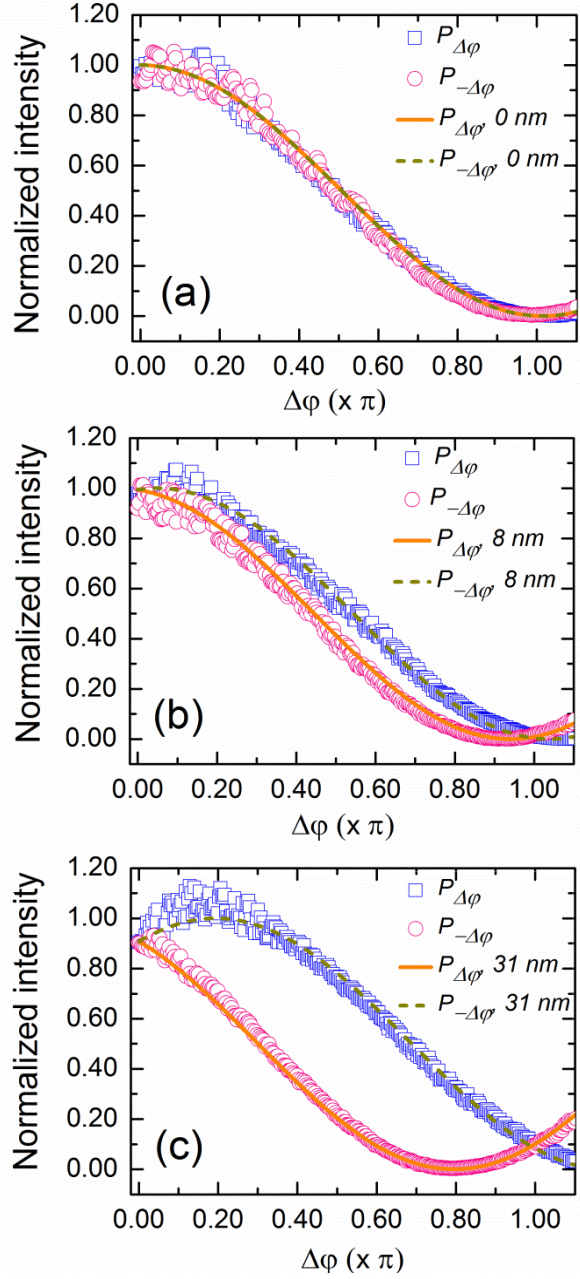

FIG 2. (Color online) Normalized intensities for $P$ when projected onto a mode of phases $\Delta\varphi$ and $-\Delta\varphi$ for different heights: (a) 1.9 nm for sample 1, (b) 9.7 nm for sample 2 and (c) 29.0 for sample 3. All measurements have standard error of 0.2 nm. The measurements based on the experimental data is discussed in the text. Theoretical curves where calculated based on the measurement done with a commercial profilometer.

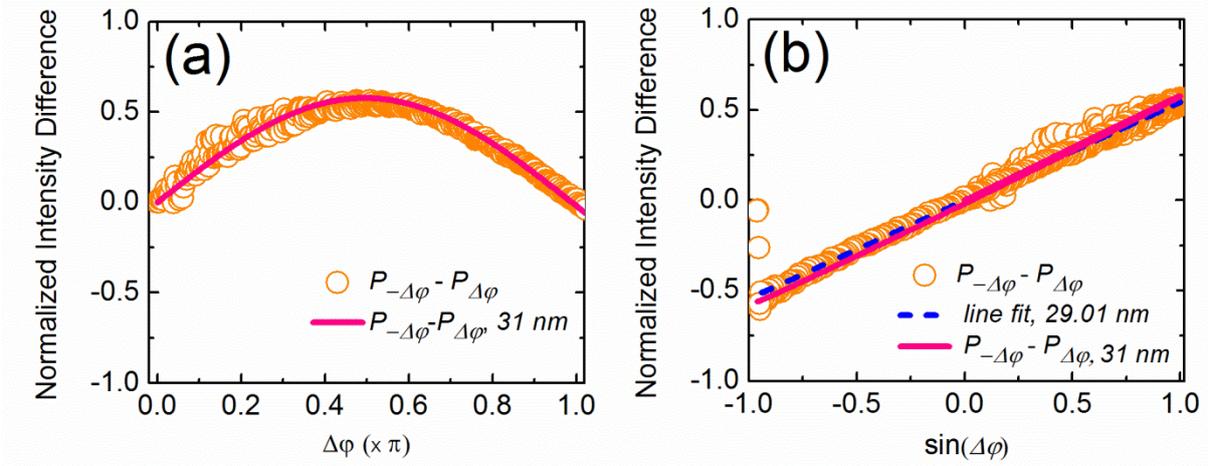

FIG. 3. (Color online) Typical data for analysis. (a) Normalized power difference $P_{\Delta\varphi} - P_{-\Delta\varphi}$ as a function of $\Delta\varphi$. (b) The difference as a function of $\sin(\Delta\varphi)$ is linear as described by Eq. (4). Line fit is from the calculated height (dashed line) and from theoretical calculations (solid line). For all plots, the theoretical curve is calculated from a cliff height of 31 nm which is independently measured with a profilometer.

Table I. Experimental thickness layer (in nanometers)

| | Profilometry[a] | CPI with mode projection[b] | Std Err |
|---|---|---|---|
| *Sample 1* | 0 | 1.89 | 0.23 |
| *Sample 2* | 8 | 9.72 | 0.17 |
| *Sample 3* | 31 | 29.01 | 0.21 |

[a]Alpha-Step IQ Surface profilometer. The standard error in the height obtained with the profilometer is in order of nanometers (~1-2 nms) owing to the uneven surface of the sample.

[b]The offset of ~ 1.89 nm is due to the nonlinear relationship between the gray level value of an image and the phase that the SLM renders.

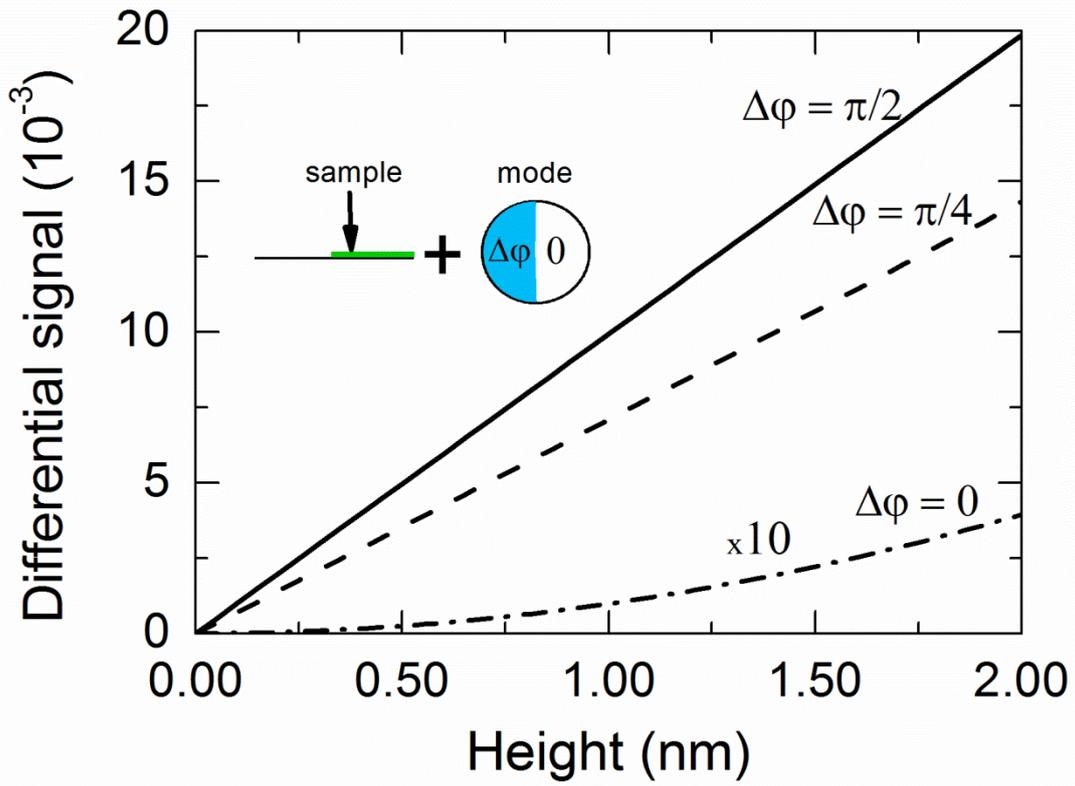

FIG 4. (Color online) Normalized differential signal $\beta_1 - \beta_2$ as a function of the sample height when the reflected signal is projected onto a Gaussian mode with phase step $\Delta\varphi$. The solid, dash and dash-dotted ($\times 10$) lines correspond to $\Delta\varphi = \pi/2, \pi/4, 0$, respectively.